# Theoretical insights into the chemical bonding, electronic structure, and spectroscopic properties of the lanarkite Pb₂SO₅ structure

Guilherme S. L. Fabris,[a] Mateus M. Ferrer,[a] Julio R. Sambrano,[b] Cláudio R. R. Almeida,[c] Carlos A. Paskocimas[c,d] and Felipe A. La Porta[e,f]*

[a] Graduate Program in Materials Science and Engineering, Technological Development Center, Universidade Federal de Pelotas, 96010-610, Pelotas, RS, Brazil

[b] Modeling and Molecular Simulation Group, São Paulo State University, Bauru, São Paulo, 17033-360, Brazil

[c] Department of Materials Engineering, Federal University of Rio Grande do Norte, 59078-970 Natal, RN, Brazil

[d] Postgraduate Program in Science and Engineering of Materials, Federal University of Rio Grande do Norte, 59078-970 Natal, RN, Brazil

[e] Federal University of Technology - Paraná, Nanotechnology and Computational Chemistry Laboratory, Avenida dos Pioneiros 3131, CEP 86036-370, Londrina - PR, Brazil

[f] State University of Londrina, Post-Graduation Program in Chemistry, Rodovia Celso Garcia Cid, 445, km 380, CEP 86057-970, Londrina - PR, Brazil

*Email: felipelaporta@utfpr.edu.br

## Abstract

Lanarkite is a mineral formed by a combination of lead, sulphur, and oxygen atoms arranged in the general chemical formula Pb₂SO₅ (PSO) that crystallises with monoclinic symmetry (belonging to the C2/m space group, No. 12). This mineral was first discovered in Lanarkshire, Scotland and was named after its location. PSO has a unique structure comprising alternating penta-coordinated lead [PbO₅] and tetra-coordinated sulphur [SO₄] clusters. This lanarkite-type structure has recently attracted significant scientific interest and has been the focus of the superconducting material research community. However, its chemistry needs to be explored further. This article presents a comprehensive investigation on the chemical bonding, electronic structure, and spectroscopic properties of the lanarkite-type PSO structure from a computational perspective. Thus, different functionals in the DFT (e.g., PBE, PBE0, PBESOL, PBESOL0, BLYP, WC1LYP37, and B3LYP) were assessed to accurately predict their fundamental properties. All the DFT calculations were performed using a triple-zeta valence plus polarisation basis set. Among all the DFT functionals tested in this study, PBE showed the best agreement with the experimental data available in the literature. Our results also reveal that the [PbO₅] clusters are formed with three Pb–O bond lengths, with values of about 2.32, 2.59, and 2.84 Å, respectively, while the [SO₄] clusters have the same S–O bond length of 1.57 Å. We performed a complete topological analysis of this system to comprehend these structural differences. Additionally, the PSO structure has an indirect band gap energy of 2.9 eV and an effective mass ratio ($m_h^*/m_e^*$) of about 0.415 (using PBE calculations) which may, in principle, indicate a low recombination of electron-hole pairs in the lanarkite structure. Therefore, we believe that a detailed understanding of their electronic structures, spectroscopic properties as well as their chemical bonding is critically important for developing new technologies based on PSO.

## 1. Introduction

In materials science, the quest for novel advanced materials, including nanostructures, with highly tailored properties and unique functionality has always been closely related to technological advancement. From this perspective, the discovery/design, and optimization of such materials can currently be accessible by different synthetic routes, drastically influencing their outstanding properties.[1–5] Notably, these advanced materials have, in the past, been designed and manufactured empirically, primarily through numerous trial and error using a combinatorial synthesis strategy.[6–10] In general, it is known that this traditional path involves high costs and takes a lot of time, in addition to the high number of possible combinations. However, the method of designing and discovering new advanced materials has changed significantly in recent decades through the combined use of theoretical/experimental strategies, which in turn enable a deeper understanding of the relationship between the structure-composition-property of these new materials.[3,9,11–16]

Therefore, new emerging technologies currently focus on the design process of new materials based on classes little explored in the literature.[17–20] In this direction, more recently, researchers now find that the combination of the lanarkite-type $Pb_2SO_5$ (abbreviated as PSO) structure with copper phosphide ($Cu_3P$) leads to the formation of $Pb_{10-x}Cu_x(PO_4)_6O$ which can contain a room-temperature superconducting state.[21,22] Such results have been generating great debates and new studies are, in turn, already calling that discovery into question.[23–27] Thus, from this perspective, PSO is now at the centre of attention of the scientific community.

As is commonly recognised, the PSO structure generally crystallizes in the monoclinic phase belonging to C2/m space group (no 12).[28,29] The initial structural description of these ceramic materials was provided by Schrauf in 1877.[30] Despite that experimental and theoretical studies involving PSO-based material applications are relatively scarce in the literature. Hence, these new advanced materials can exhibit a plethora of attractive physical and chemical properties, that need to be discovered as well as exploited correctly for their prowess commercialization in the future. We believe that a detailed understanding of their electronic structure and spectroscopic properties as well as their chemical bonding is critically important for developing new technologies based on PSO.

In this paper, we provide a comprehensive perspective on the chemical bonding, electronic structure, and spectroscopic properties of the lanarkite-type PSO structure from a computational perspective, through the Density Functional Theory (DFT), where it is investigated the influence of seven different functional over the structural and electronic properties, to unravel who best describes its properties. In this case, the proposed computational strategy is of large interesting in order to a more accurate description of electronic parameters, which govern much of their technological applications.

## 2. Computational Method and Model System

All the periodic DFT calculations have been performed using seven density functionals, PBE,[31] PBE0,[32] PBESOL,[33] PBESOL0, BLYP,[34,35] WC1LYP[36] and B3LYP,[34,35,37] for comparison as implemented in CRYSTAL17[38] software. The Pb, S and O atomic centers were described by a triple-zeta valence plus polarization (TZVP).[39] The minimum energy structure was confirmed by diagonalizing the Hessian matrix with respect to the lattice parameters and internal coordinates, followed by the confirmation of the absence of imaginary frequencies. The structural convergence was checked on the gradient components and nuclear displacements, with tolerances on their root mean square set to 0.0001 and 0.0004 a.u., respectively. The precision of the infinite Coulomb and HF exchange series is controlled by five parameters $\alpha_i$, with $i = 1-5$, such that two-electron contributions are neglected when the overlap between atomic functions is lower than $10^{-\alpha_i}$. In our simulation, the five $\alpha_i$ parameters have been set to 8, 8, 8, 8, and 18 and the shrinking factor for the Pack-Monkhorst and Gilat nets was set to 9. The band structure and density of states (DOS) were analysed using the Properties17 routine implemented in the CRYSTAL code, with the same $k$-point sampling employed for the diagonalization of the Fock matrix in the optimization process, for the DOS, and the high-symmetry points in the Brillouin zone, for the band structure. Those vibrational frequencies were determined at the $\Gamma$ point using the numerical second derivatives of the total energies, which was estimated using the coupled perturbed Hartree-Fock/Kohn-Sham algorithm.[40–42] The chemical bonds and electron density distribution properties were investigated applying the Quantum Theory of Atoms in Molecules and Crystal (QTAIMAC),[43] with Topond program,[44] as implemented in CRYSTAL17.

### 3. Results and Discussion

#### a. Structural properties

Figure 1(a) shows the representation of the PSO unit cell, with experimental lattice parameters of $a = 13.746$ Å, $b = 5.696$ Å and $c = 7.066$ Å, and angles $\alpha = \gamma = 90°$ and $\beta = 115.79°$.[45] The lanarkite PSO structure has monoclinic symmetry (belonging to the C2/m space group) and is composed of a combination of altered penta-coordinated lead [PbO$_5$] and tetra-coordinated sulphur [SO$_4$] clusters. Figures 1(b) and 1(c) show the results of our investigation of the influence density functional on the lattice parameters. The dashed lines indicate the experimental values for the respective parameters.

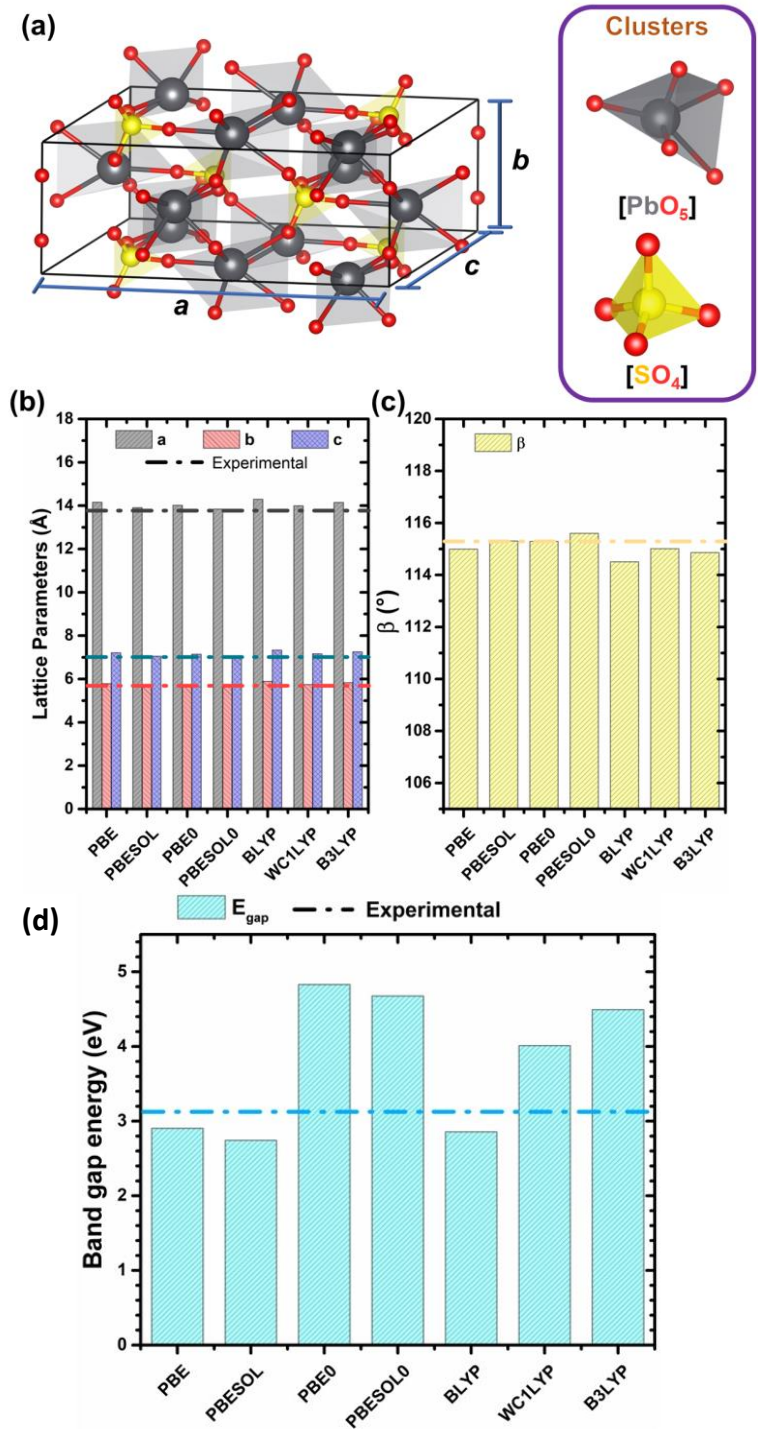

**Figure 1**: (a) Bulk structure unit cell and its constituting clusters, (b) lattice parameters, (c) β angle and (d) band gap energy variations in function of the density functional. The dash dotted line represents the reference values for each parameter.[45,46]

As shown in Figures 1(b) and 1(c), all density functionals used in this study can accurately describe the PSO lattice, with small variations observed between the theoretical and experimental values. For instance, the sum of the deviations found for all these lattice parameters (in percentage) was for PBESOL0 (0.71%), PBE0 (0.67%), PBESOL (0.21%), and PBESOL0 (0.16%). However, it is well known that in the development of new semiconductor materials, the band gap is the most fundamental property.[3] Among all the density functionals tested in this study, PBE presented a band gap value closest to that reported in Materials Project database,[46] as can be seen in Figure 1(d), with a 7.20% deviation. Therefore, all subsequent analyses of this system were performed using the PBE functional.

The PBE calculated lattice parameters of our structure is $a$ = 14.16Å (-2.99%), $b$ = 5.78Å (-1.54%), and $c$ = 7.21Å (-2.02%), with an angle β of 115.0° (0.68%); which are in good agreement with most of the experimental studies.[45] In this lanarkite structure, the [PbO$_5$] clusters are formed with three Pb–O bond lengths, with values of approximately 2.32, 2.59, and 2.84 Å, respectively, while the [SO$_4$] clusters have the same S–O bond length of 1.57 Å. Therefore, we believe that these distortions in the [PbO$_5$] clusters are the origin of their physical properties.

This study also evaluated the spectroscopic properties of the optimised structure calculated at PBE level through Raman spectroscopy and XRD, as they provide chemical and structural fingerprints for the monoclinic PSO structure. Figures 2(a) and 2(b) show the XRD patterns and the simulated Raman spectra of the PSO structure. Figure 2(a) compares the theoretical and experimental PSO structured XRD patterns. Furthermore, the results are similar to those obtained by experimentation, and are more pronounced at the lower angles. Notably, in this study the experimental PSO sample were obtained using the solid-state reaction method at 725 ºC for 24 h, which will be the target of future studies. Furthermore, the PSO has 24 active Raman modes, of which only six are the most relevant and intense, as shown in Figure 2(b). These intense modes are located at 142.95 cm$^{-1}$ (A$_g$, high lead rocking and small sulphur wagging), 344.85 cm$^{-1}$ (B$_g$, lead rocking and sulphur scissoring), 357.11 cm$^{-1}$ (A$_g$, lead and sulphur symmetrical stretching), 844.55 cm$^{-1}$ (A$_g$, sulphur symmetrical stretching), 957.91 cm$^{-1}$ (B$_g$, sulphur asymmetrical stretching) and 974.43 cm$^{-1}$ (A$_g$, sulphur asymmetrical stretching and wagging). A video animation of the vibrational Raman-active mode movements is provided in Supplementary Material.

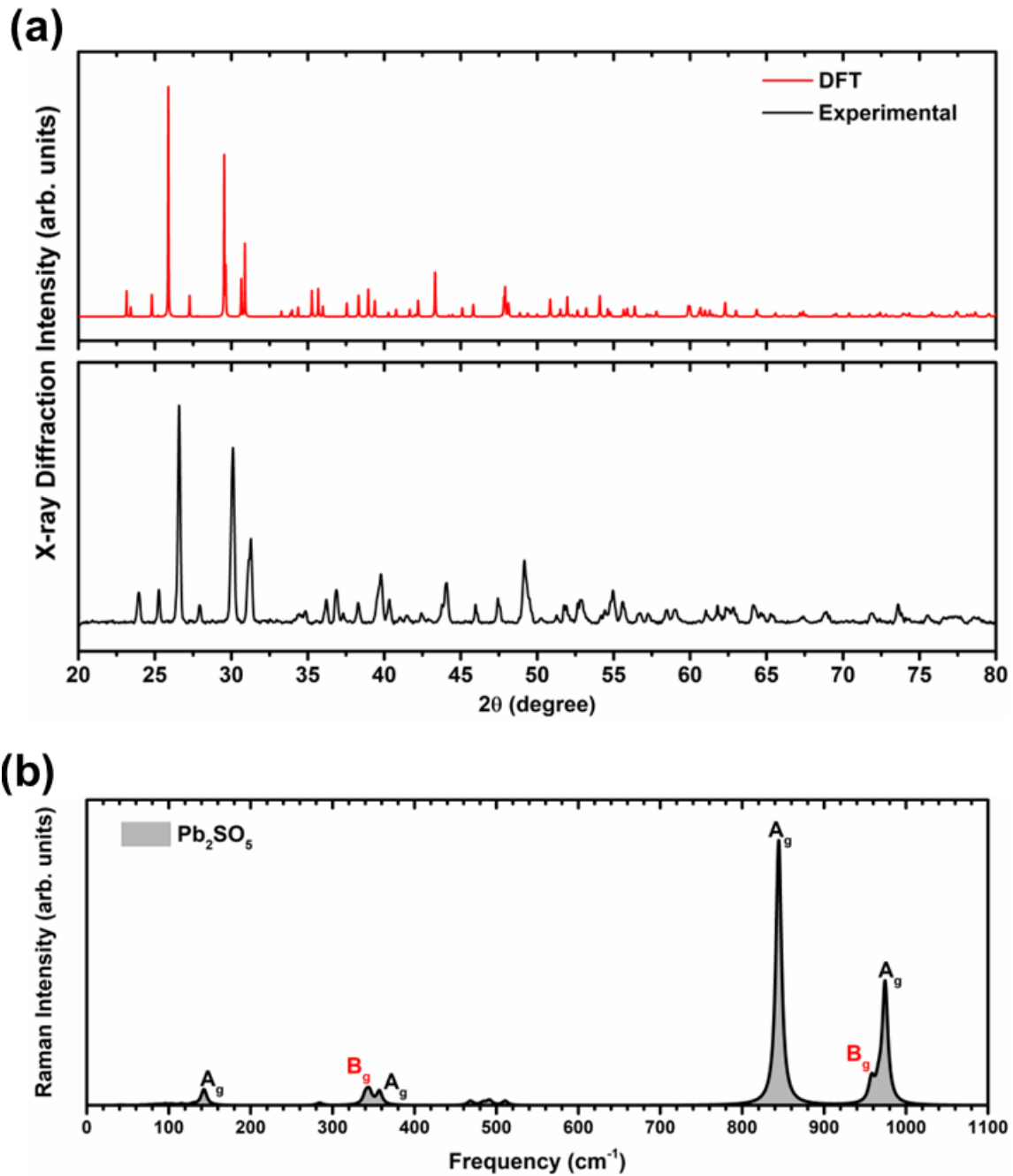

**Figure 2**: (a) X-ray diffraction curves (theoretical and our own experimental) and (b) Theoretical vibrational Raman spectra and of PSO structure.

The observed lower-frequency vibrational modes can be attributed to a combination of the vibrations of the [PbO$_5$] and [SO$_4$] clusters, with a dominant vibration from the [PbO$_5$]; whereas, at higher frequencies, we observed that [SO$_4$] is the only cluster that vibrates.

### b. Topological properties

To understand the nature of bonding in the PSO structure, the topological properties were investigated using QTAIMAC analysis at the bond critical points (BCP). Table 1 shows the topological parameters for four different BCPs found in the [PbO$_5$] cluster and two in the [SO$_4$] cluster.

**Table 1:** Topological properties of PSO measured at the BCPs: electron density ($\rho(r)$), Laplacian of electron density ($\nabla^2\rho(r)$), $|V|/G$ ratio, bond degree ($H/\rho(r)$), ellipticity ($\varepsilon$), and bond type. Labels "cov", "T", and "ionic" denote covalent, transitory, and ionic bond (interactions) types, respectively.

| Bond Critical Points | | | | | | | | |
| --- | --- | --- | --- | --- | --- | --- | --- | --- |
| **A-B** | **ρ** | **∇²ρ** | **G** | **V** | **\|V\|/G** | **H/ρ(r)** | **ε** | **bond** |
| Pb-O$_1$ | 0,065 | 0,209 | 0,063 | -0,073 | 1,168 | -0,163 | 0,011 | T cov |
| Pb-O$_2$ | 0,031 | 0,112 | 0,028 | -0,028 | 0,999 | 0,001 | 0,010 | T |
| Pb-O$_3$ | 0,020 | 0,065 | 0,015 | -0,014 | 0,927 | 0,055 | 0,075 | ionic |
| Pb-O$_4$ | 0,012 | 0,038 | 0,008 | -0,007 | 0,852 | 0,100 | 0,100 | ionic |
| S-O$_1$ | 0,224 | 0,278 | 0,289 | -0,509 | 1,760 | -0,983 | 0,001 | T cov |
| S-O$_2$ | 0,224 | 0,251 | 0,284 | -0,506 | 1,779 | -0,988 | 0,006 | T cov |

The [PbO$_5$] cluster BCPs are arranged in a crescent order of bond length, which shows that the smaller bond distances have a greater concentration of electron density at the BCP and exhibit a transitory incipient covalent bond character owing to the negative covalent degree. As the bond distance increases, the incipient covalent starts to change to a more directional form. Consequently, a transitory ionic behaviour is induced, which is, of course, confirmed by the small concentration of electron density at this point, and the bond degree ($H/\rho(r)$) is close to 0. At greater bond distances the bonding changes to completely ionic, where the electron density is almost completely at the atoms and not at the BCP; however, the ellipticity suggests a small covalent behaviour (no such spherical shape of the bond). For [SO$_4$], the bonds have a high concentration of charge density, a high $|V|/G$ ratio, and negative bond degree, indicating an incipient covalent bond with spherical bond shape. To generate a three-dimensional visualization of the QTAIMAC descriptors, in this study was used TopIso3D Viewer [47]. These results can be seen in the HTML file in the Supplementary Material.

Additionally, Mulliken analysis showed that in the [$PbO_5$] cluster, the $Pb - O$ bonds had an overlap population of 64, 40, and 34 m|e|, whereas in the [$SO_4$] clusters the S–O bonds had a bond overlap of about 528 m|e|. This result suggests that the sharing of electrons is more in the [$SO_4$] cluster than in the [$PbO_5$], as expected.

### c.   Electronic properties

The band gap energy (see Figure 1(d)) was also compared with reported experimental data of approximately 3.13 eV [46] and is best described by PBE followed by the BLYP functional. Figure 3(a) displays the band structure and DOS of the monoclinic PSO structure obtained through the high-symmetry path ($\Gamma$—C|$C_2$—$Y_2$—$\Gamma$—$M_2$—D|$D_2$—A—$\Gamma$|$L_2$—$\Gamma$—$V_2$) along the Brillouin zone. The analysis of PBE calculations shows that the band are well-defined, and it has an indirect band gap energy of 2.9 eV, where the valence band (VB) maximum is located at C point, while the conduction band (CB) minimum is at $M_2$ (Figure 4(a)). The projected DOS (Figure 3(a)) shows that the O states have a major influence around the band gap, with the $2p_x$, $2p_y$, and $2p_z$ orbitals, and a minor influence of the Pb 6s, $6p_y$, and $6p_z$ states at the VB, with no presence of S states around the band gap. In the CB, the major influence is from the Pb $6p_x$, $2p_y$, and $2p_z$ orbitals, and a minor influence is from the O $2p_z$, S 2s, and $2p_x$ orbitals. Interestingly, S atoms have a greater influence at lower and higher energy values, far from the forbidden zone. For clarity, Figure S1 (see Supplementary Material) shows the DOS as a function of the elemental orbital.

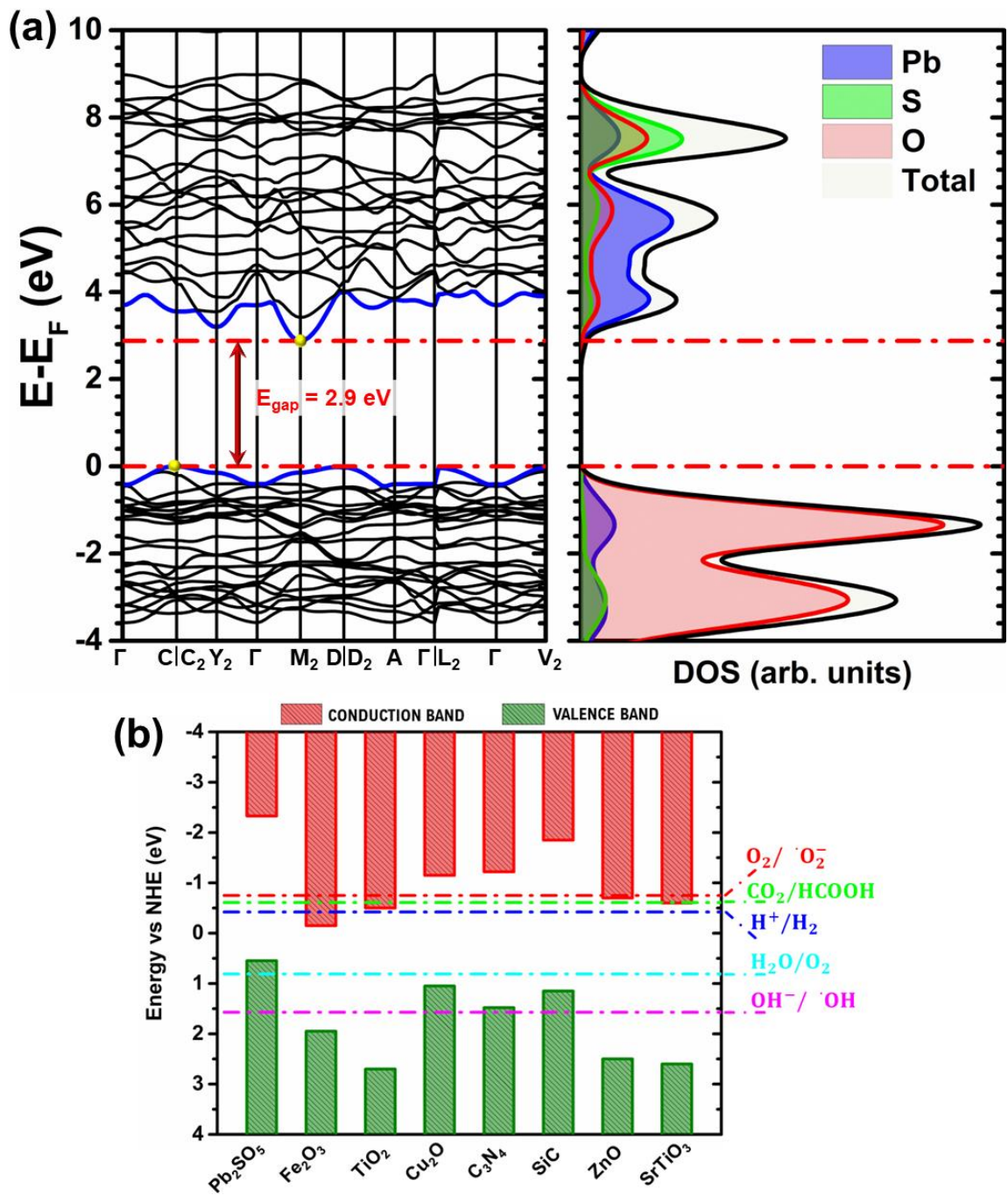

**Figure 3**: (a) Band structure of PSO in the high-symmetry path at the Brillouin zone and its density of states (DOS), and (b) Band edge positions for PSO (our work) and for other materials.[48,49]

The band edge positions (Figure 3(b)) was also elaborated to analyse the potential use of PSO, along with other materials [48,49], in the photocatalytic water separation using solar energy to produce oxygen ($O_2$) and hydrogen ($H_2$), as well as to photodegrade pollutants via reduced oxygen species (ROS). For a semiconductor photocatalyst to work well, it needs to have the right energy levels of its CB and VB so that they can match the energy needed to split water into hydrogen and oxygen. This property can be evaluated through the redox potentials $E_0(O_2/\ ^\bullet O_2^-)$ and $E_0(OH^-/\ ^\bullet OH)$, which was highlighted to evaluate the possibility of ROS generation. As can be seen in Figure 3(b), the PSO, which is a p-type semiconductor, have a good band edge positioning for the reduction processes of $CO_2$, $O_2/\ ^\bullet O_2^-$ and $H^+/H_2$, however it is not favourable for the occurrence of the oxidation process $H_2O/O_2$ and $OH^-/\ ^\bullet OH$, as indicated by the dashed dotted lines.

To understand better the mobility of electrons in the bands, the effective mass of electrons ($m_e^*$) and holes ($m_h^*$) were determined, and they are inversely proportional to the second derivative of the band energy with respect to the k-point, which implies that the greater the effective mass, the more difficult it is to drive them through the bands. [50,51] Hence, the effective mass tensor definition is given by:

$$\left(\frac{1}{m^*}\right)_{\mu\nu} = \frac{1}{\hbar}\frac{d^2 E_n(k)}{dk_\mu dk_\nu}\ , \qquad (1)$$

where $E_n(k)$ is the n-th band energy, and $k_\mu$ and $k_\nu$ are the k-points in the Brillouin zone at the $\mu$ and $\nu$ directions, respectively.[50,51] This indicates that the hole-effective mass is heavier than the electron-effective mass because the lowest energy CB has a parabolic shape, as indicated by the yellow point in the Figure 3(a). For the effective mass analysis, a symmetrical path in the Brillouin zone of $\Gamma$-C-$\Gamma$ (VB) and $\Gamma$-$M_2$-$\Gamma$ (CB) was used. PBE calculations showed that PSO has an effective mass of electrons ($m_e^*/m_0$) and holes ($m_h^*/m_0$) and mass ratios ($m_h^*/m_e^*$) of about 5.610, 2.328, and 0.415, respectively. As the $m_h^*/m_e^*$ ratio is small, we have a lower recombination ratio of electron-hole pairs, which can, in principle, be interesting for p-type conductivity [52].

**Conclusion**

In summary, we have provided very rich computational information to better understand of the physical and chemical properties of the monoclinic PSO structure. For bulk calculations, different functionals in the periodic DFT (e.g., PBE, PBE0, PBESOL, PBESOL0, BLYP, WC1LYP37, and B3LYP) were assessed to accurately predict their properties. All the density functionals employed in this study have proven to be suitable for accurately reproducing the experimental lattice parameters of the PSO. Notably, PBE exhibited the closest agreement with the experimental band gap values. Our DFT calculations show that the PSO is a p-type semiconductor with an indirect band gap of about 2.9 eV and an effective mass ratio ($m_h^*/m_e^*$) of about 0.415 (using PBE calculations). These theoretical findings open new opportunities in research with lanarkite-type materials, and hence are of great importance to describing their exotic functional properties, revealing an in-depth understating on the nature of their chemical bonds, as well.

## Acknowledgements

GSL Fabris thanks the postdoc scholarship financed by National Council for Scientific and Technological Development – CNPq (150187/2023-8) and Fundação de Amparo à Pesquisa do Estado do Rio Grande do Sul – FAPERGS. C. A. Paskocimas acknowledges CNPq for financial support through grant #482473/2010-0, #446126/2014-4, #308548/2014-0 and #307236/2018-8. The computational facilities were supported by resources supplied by the Molecular Simulations Laboratory (São Paulo State University, Bauru, Brazil).

# Supplementar Material

# Theoretical insights on the chemical bonding, electronic structure and spectroscopic properties of the lanarkite Pb₂SO₅ structure

Guilherme S. L. Fabris,[a] Mateus M. Ferrer,[a] Julio R. Sambrano,[b] Carlos A. Paskocimas,[c] Cláudio R. R. Almeida[d] and Felipe A. La Porta[e]*

[a] *Graduate Program in Materials Science and Engineering, Technological Development Center, Universidade Federal de Pelotas, 96010-610, Pelotas, RS, Brazil*

[b] *Modeling and Molecular Simulation Group, São Paulo State University, Bauru, São Paulo, 17033-360, Brazil*

[c] *Postgraduate Program in Science and Engineering of Materials, Federal University of Rio Grande do Norte, 59078-970 Natal, RN, Brazil*

[d] *Department of Materials Engineering, Federal University of Rio Grande do Norte, 59078-970 Natal, RN, Brazil*

[e] *Federal Technological University of Paraná, UTFPR, Avenida dos Pioneiros 3131, Londrina 86036-370, PR, Brazil*

**\*Email:** felipelaporta@utfpr.edu.br

In Table S1 it is shown the data of the structural and electronic parameters in function of the density functional.

**Table S1**: Influence of the functional in the lattice parameters (a, b and c in Å; and β in degrees) and band gap energy ($E_{gap}$, in eV). The Δ indicates the percentual difference from the experimental to the calculated data.

|          | a      | Δa     | b     | Δb     | c     | Δc     | β      | Δβ    | $E_{gap}$ | $\Delta E_{gap}$ |
|----------|--------|--------|-------|--------|-------|--------|--------|-------|-----------|------------------|
| **PBE**  | 14,16  | -2,99% | 5,78  | -1,54% | 7,21  | -2,02% | 115,00 | 0,68% | 2,90      | 7,23%            |
| **PBESOL** | 13,91 | -1,17% | 5,66  | 0,69%  | 7,05  | 0,21%  | 115,31 | 0,42% | 2,74      | 12,32%           |
| **PBE0** | 14,02  | -1,97% | 5,73  | -0,67% | 7,14  | -1,08% | 115,29 | 0,43% | 4,83      | -54,31%          |
| **PBESOL0** | 13,84 | -0,71% | 5,64 | 0,91%  | 7,03  | 0,47%  | 115,60 | 0,16% | 4,68      | -49,41%          |
| **BLYP** | 14,29  | -3,98% | 5,89  | -3,39% | 7,33  | -3,78% | 114,51 | 1,11% | 2,86      | 8,73%            |
| **WC1LYP** | 13,99 | -1,78% | 5,74  | -0,85% | 7,17  | -1,42% | 115,01 | 0,67% | 4,01      | -28,18%          |
| **B3LYP** | 14,15 | -2,91% | 5,83  | -2,31% | 7,26  | -2,69% | 114,86 | 0,80% | 4,49      | -43,56%          |
| **Exp.** | 13,746 | -      | 5,696 | -      | 7,066 | -      | 115,79 | -     | 3,13      | -                |

The orbital contribution of each element present in the lanarkite PSO can be seen in Figure S1 as shown below.

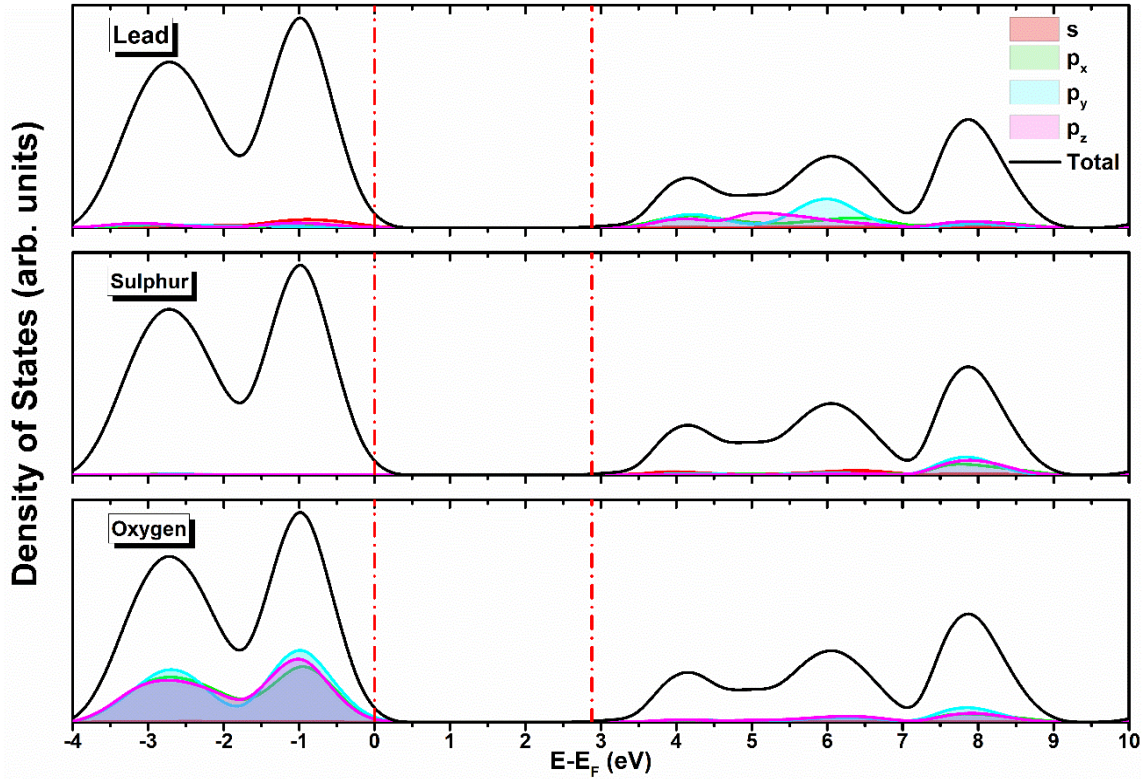

**Figure S1**: Partial Density of States of lanarkite PSO.